\documentclass[aps,pre,twocolumn,groupedaddress,showkeys]{revtex4-1}
\usepackage{graphicx}
\usepackage{epstopdf}
\usepackage{amsmath}
\begin{document}


\title{Semi-analytic path integral solution of SABR and Heston equations: 
pricing Vanilla and Asian options}


\author{Jan Kuklinski}
\author{Kevin Tyloo}

\affiliation{Facult\'e des HEC, Universit\'e de Lausanne, CH-1015 Lausanne, Switzerland}


\date{\today}

\begin{abstract}

\noindent We discuss a semi-analytical method for solving SABR-type equations based on path integrals. In this approach, one set of variables is integrated analytically while the second set is integrated numerically via Monte-Carlo.  This method, known in the literature as Conditional Monte-Carlo, leads to compact expressions functional on three correlated stochastic variables. The methodology is practical and efficient when solving Vanilla pricing in the SABR, Heston and Bates models with time depending parameters. Further, it can also be practically applied to pricing Asian options in the $\beta=0$  SABR model and to other $\beta=0$ type models.

\end{abstract}


\maketitle




\section{Introduction}

The SABR model introduced by Hagan et al.  \cite{HagLes,JanObl} is a canonical Stochastic Volatility model: 

\begin{eqnarray}
d\tilde{S}_{t} &=& S_{0}\Big(\frac{\tilde{S}_{t}}{S_{0}}\Big)^{\beta}\tilde{\sigma}_{t}\{\rho d\tilde{V}_{t}+\sqrt{1-\rho^{2}}d\tilde{W}_{t}\}\\
d\tilde{\sigma}_{t} &=& \nu\tilde{\sigma}_{t}d\tilde{V}_{t}
\end{eqnarray}

This framework provides a natural extension of the Bachelier/Normal, Black-Scholes models and of the Shifted Log-Normal equation. As discussed in \cite{JRK_D,Lor1}, setting $\beta=0$ and $\rho=\pm1$ leads to the Shifted Log-Normal model. The very important characteristic of the SABR dynamics is the linearity in the $\tilde{W}_{t}$ stochastic process. This follows the structure of the Heston equations \cite{HestMod} and of the Stein and Stein model \cite{SteinStein} established earlier.

The Authors of the SABR model provided very accurate pricing of Vanilla options using approximated/asymptotic solutions \cite{HagLes}. These solutions are useful in most applications, yet in selected problems a higher accuracy is needed. The concept of analytical solutions in the form of multiple integrals was discussed in the literature \cite{HenLab} and finally mastered for the $\beta=0$  SABR by Korn et al \cite{Kor}.

Unlike for the Shifted-Log-Normal model, these `base' solutions are not providing explicit analytic expressions for the stochastic trajectories, which are needed for path dependent contracts such as Asian options and Structured Products. What are also missing are generalizations and solutions of the   $\beta=0$ SABR equations to a mean-reverting case.

The target of this work is twofold. First we use path-integral solutions of SABR equations with time-dependent parameters to price Vanilla options ($\beta=0$ \textit{or} $1$) together with $\beta=0$ Asian options. We show that option prices are functional on three correlated stochastic variables. Second we discuss applying the same methodology to mean reverting models, while for the $\beta=1$ SABR there is the Heston and Sch\"obel-Zhu \cite{SchZhu} models.
   
\begin{figure}
\includegraphics[width=\columnwidth]{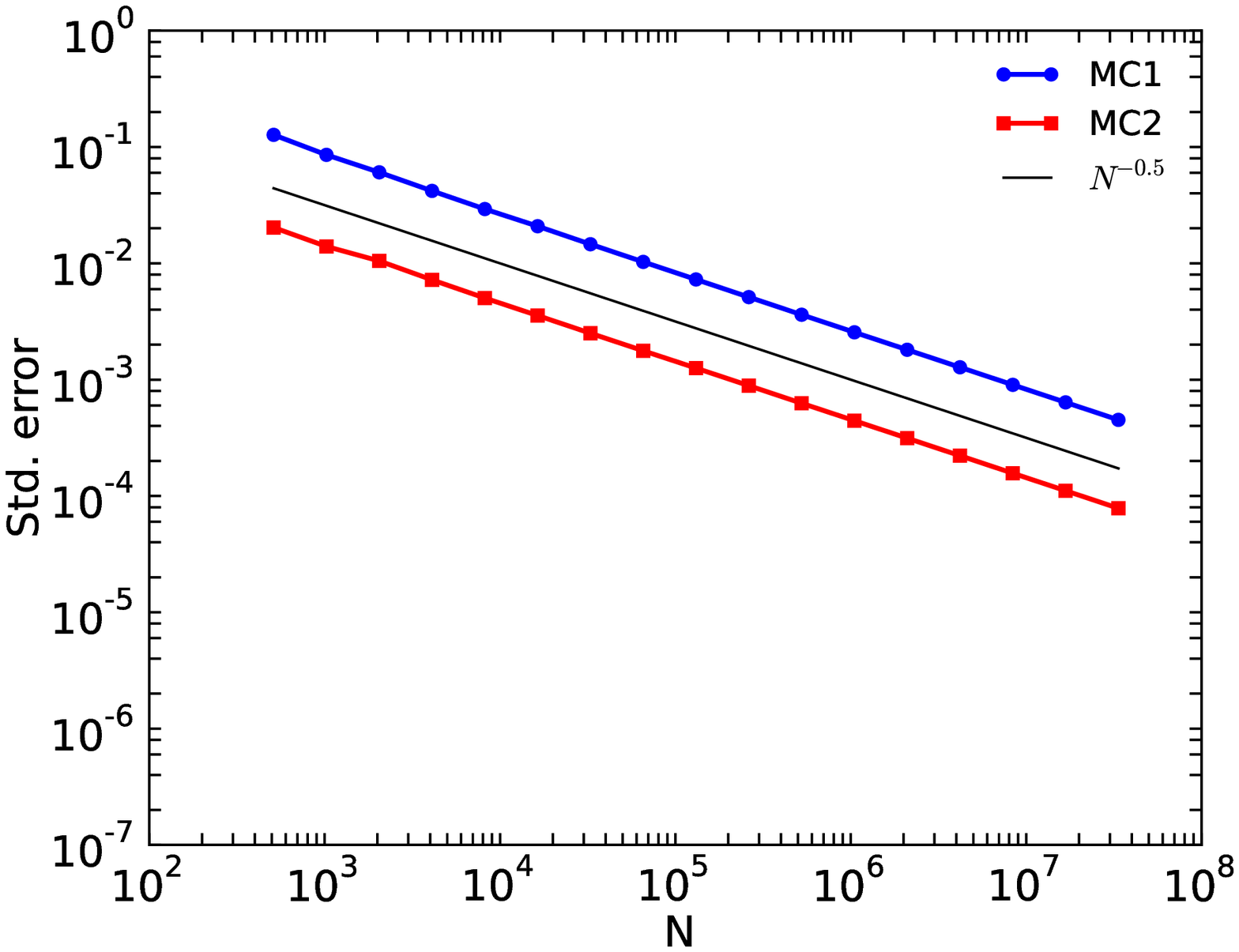}\llap{\makebox[.84\columnwidth][l]{\raisebox{1cm}{\includegraphics[width=.44\columnwidth]{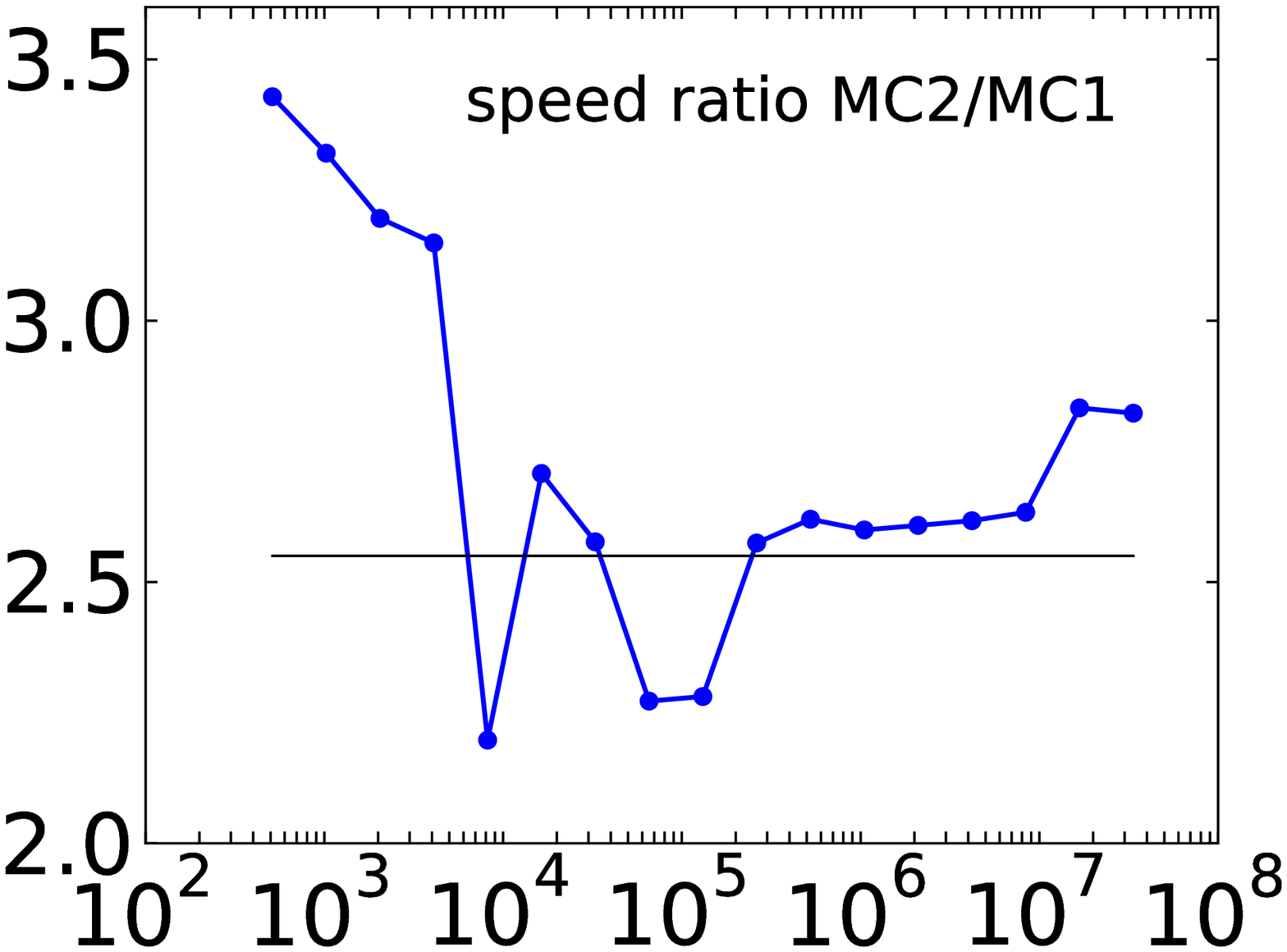}}}}
\caption{\label{conv}Comparison of the convergence between Monte-Carlo 1 (MC1) and 2 (MC2) for Vanilla option. \textit{Inset:} MC2 is about twice as fast as MC1.}
\end{figure}

We stress that the methodology we are using requires only one dimension to be Monte-Carlo integrated as the second one can be integrated analytically. Such semi-analytic integration technique (we call it also MC2 integration), is known in the literature as Conditional Monte-Carlo \cite{Will}. We found the term semi-analytical Monte-Carlo integration (also referred as MC2) useful and explanatory as it indicates diminishing the numerical task by analytical means.  The computational task is twice smaller when the MC2 method is used. And the error, although still converging in square-root, is smaller than the classical approach even for a few paths (see FIG.\ref{conv}). It is also interesting to note that strongly out-of-the-money options can be priced without having to simulate very large number of paths.

The issue of pricing arithmetic Asian options is very widely discussed in the literature. Many authors have tried to solve or approximate the pricing of the arithmetic Asian options, including solving numerically the Heston dynamics \cite{Print}\footnote{An extensive review of the pricing od Asian options is given in \cite{Lor}.}.  Yet the greatest effort has been put on finding or approximating the distribution of a sum of log-normally distributed random variables. This sum arises naturally for arithmetic Asian options under the Black-Scholes model due to the average in the payoff. Milevsky and Posner \cite{MilPos} approximates the finite sum of Lognormals by an inverse Gamma distribution which is the true density if the sum was infinite \cite{Duf,Yor, Yor3}. A similar idea was applied by Levy \cite{Lev} and Turnbull and Wakeman \cite{TurWak} but using a log-normal distribution  \footnote{The sum of correlated lognormals is also of important topics in other fields such as telecommunication.}. However the efforts done for the Black-Scholes model cannot account for market skew and smile. 
    
In contrast to the rather complex calculations made for Asian options in the Black-Scholes model, the semi-analytic the $\beta=0$ SABR pricing for Asian options presented herein is very simple and efficient and to the best of our knowledge was not discussed prior in the literature. It should be stressed that the methods applied to other $\beta=0$ models and that such framework can be considered as generalizations of the Heston, Bates \cite{Bat} and Sch\"obel-Zhu models. In contrast to Vanilla pricing, the Asian options pricing simplicity cannot be reproduced outside the SABR $\beta=0$  model.
    
For Vanilla pricing, the semi-analytic MC2 method can be applied for both $\beta=0$  and $\beta=1$. This type of calculation can be used to check the accuracy of the Hagan solutions for the SABR model. It is worth stressing that for $\beta=1$, solutions discussed by Korn et al. are to the best of our knowledge not available. Most importantly, the MC2 calculation provides a reliable solution to stochastic equations with time dependent parameters. Last but not least it can also be used as a comparative tool to calibrate general Monte-Carlo pricing scheme.
    
The presented calculation procedure is that it can be extended to mean-reverting models including jumps. We briefly discuss its application to the Heston, Bates and Sch\"obel-Zhu models.

\section{Path integral solution of the $\beta=1$ and $\beta=0$ SABR equations}

We start with $\beta=1$ SABR model:

\begin{align}
d\tilde{S}_{t} &= \tilde{S}_{t}\tilde{\sigma}_{t}\{\rho d\tilde{V}_{t} + \sqrt{1-\rho^{2}}d\tilde{W}_{t}\} \\
\tilde{\sigma}_{t} &= \sigma_{0}\exp[\nu\tilde{V}_{t}-\frac{\nu^{2}}{2}t]
\end{align}

We use the log-variable $\tilde{\Phi}_{t}$:

\begin{eqnarray}
&\tilde{\Phi}_{t}(\tilde{S}_{t},t)& = \ln[\tilde{S}_{t}/S_{0}] \\
&d\tilde{\Phi}_{t}& = -\frac{1}{2}(\tilde{\sigma}_{t})^{2}dt + \tilde{\sigma}_{t}\{\rho d\tilde{V}_{t} + \sqrt{1-\rho^{2}}d\tilde{W}_{t}\}
\end{eqnarray}

We integrate the three parts:

\begin{equation}
\tilde{\Phi}_{t} = \{-\frac{1}{2}(\tilde{f}_{t}^{x})^{2}+\tilde{f}_{t}^{y}+\tilde{f}_{t}^{z}\}
\end{equation}

Where:

\begin{align}
\tilde{f}_{t}^{x}& = \sigma_{0}\sqrt{\int_{0}^{t}e^{2\nu\tilde{V}_{s}-\nu^{2}s}ds} \\
\tilde{f}_{t}^{y}& = \frac{\sigma_{0}\rho}{\nu}\{e^{\nu\tilde{V}_{t}-\nu^{2}t/2}-1\} \\
\tilde{f}_{t}^{z}& = \sigma_{0}\sqrt{1-\rho^{2}}\int_{0}^{t}e^{\nu\tilde{V}_{s}-\nu^{2}s/2}d\tilde{W}_{s}
\label{equSABR_full}
\end{align}

We can further rewrite the variables $\tilde{f}_{t}^{z}$ using time changed integrals $\tau(t) = \nu^{2}t$ e.g.:

\begin{equation}
\tilde{f}_{t}^{z} = \frac{\sigma_{0}}{\nu}\sqrt{1-\rho^{2}}\sqrt{\int_{0}^{\vartheta=\nu^{2}t}e^{2\tilde{V}_{\tau}-\tau}d\tau}\tilde{Q}_{W}
\end{equation}

and $\tilde{Q}_{W}$ being an unit Gaussian variable ($\mathrm{E}[\tilde{Q}_{W}]=0$ and $\mathrm{E}[\tilde{Q}_{W}^{2}]=1$), uncorrelated with the Wiener process $\tilde{V}_{\tau}$.

Finally the $\beta=1$ SABR solution has the form:

\begin{equation}
\label{equSABR_b1}
\tilde{S}_{t} = S_{0}\exp[-\frac{1}{2}(\tilde{f}_{t}^{x})^{2}+\tilde{f}_{t}^{y}+\tilde{f}_{t}^{z}]
\end{equation}

The same pattern applied to the $\beta=0$ SABR:

\begin{align}
d\tilde{S}_{t} &= S_{0}\tilde{\sigma}_{t}\{\rho d\tilde{V}_{t} + \sqrt{1-\rho^{2}}d\tilde{W}_{t}\} \\
\tilde{\sigma}_{t} &= \sigma_{0}\exp[\nu\tilde{V}_{t}-\frac{\nu^{2}}{2}t]
\end{align}

leads to the following solution:

\begin{equation}
\label{equSABR_b0}
\tilde{S}_{t} = S_{0}\{1+\tilde{f}_{t}^{y}+\tilde{f}_{t}^{z}\}
\end{equation}

The formal path-integral solutions of the SABR equations as displayed in Eqs. \ref{equSABR_b1} and \ref{equSABR_b0} have a simple form depending on three parameters $\vartheta=\nu^{2}t$, $\sigma_{0}\sqrt{T}$ and $\rho$. Further for $\rho=0$ the SABR solutions are depending only on two parameters $\vartheta$ and $\bar{\sigma}=\sigma_{0}\sqrt{T}$. For $\beta=0$ and $\rho=\pm1$ we recover the Shifted Log-Normal model depending only on $\tilde{f}_{t}^{y}$. This generic form of SABR solutions was partially discussed in the literature \cite{Will,Lor, Sco}.


\section{Re-formatiing the solutions and semi-analytical integration}

When considering stochastic problems depending on a fixed time horizon $t=T$, such as Vanilla pricing, we can further simplify the structure of the solutions using another set of stochastic variables ($\vartheta=\nu^{2}T$):

\begin{align}
\tilde{y}_{\vartheta}& = \exp[\tilde{V}_{\vartheta}-\vartheta/2]-1 \\
\tilde{z}_{\vartheta}& = \sqrt{\int_{0}^{\vartheta}d\tau\exp[2\tilde{V}_{\tau}-\tau]}
\end{align}

The solutions have then the form:

\begin{align}
\tilde{f}_{t}^{x}& = \frac{\sigma_{0}}{\nu}\tilde{z}_{\vartheta} \\
\tilde{f}_{t}^{y}& = \frac{\sigma_{0}\rho}{\nu}\tilde{y}_{\vartheta} \\
\tilde{f}_{t}^{z}& = \frac{\sigma_{0}}{\nu}\sqrt{1-\rho^{2}}\tilde{z}_{\vartheta}\tilde{Q}_{W}
\label{equSABR_red}
\end{align}

The solution provided by Eq. \ref{equSABR_red} follows with the logic of the formalism introduced first by Bougerol \citep{Bou} and follows by other \cite{Yor,Yor2,Vak}.

Following Eq. \ref{equSABR_red} we see that for $t=T$ a double numerical Monte-Carlo integration over $\tilde{V}_{t}$  and $\tilde{W}_{t}$ trajectories can be replaced with a Monte-Carlo integration involving trajectories $\tilde{V}_{t}$ and a single Gaussian variable $\tilde{Q}_{W}$.

As next step is made, following Eq. \ref{equSABR_full}, for pricing problems in which the integration over the $\tilde{Q}_{W}$ variable can be performed analytically. The we generate $\tilde{V}_{t}$ trajectories via Monte-Carlo and for each volatility trajectory solve analytically the linear diffusion/pricing problem eventually averaging exact Vanilla prices over the set of trajectories. The calculation details together with numerical computations are discussed in the Appendix. We call this semi-analytic Monte-Carlo calculations, while the literature use the name of Conditional Monte-Carlo \cite{Will, Lor} or Rao-Blackwellization in the context of the Rao-Blackwell-Kolmogorov theorem \cite{Owe}.


\section{Comparing to Hagan et al asymptotic solutions} 

The MC2 solution allows evaluating the accuracy of the Hagan et al approximated solution. The control parameter of this approximation is:

\begin{equation}
\vartheta = \int_{0}^{T}\nu_{t}^{2}dt
\end{equation}

\begin{figure*}
\centering
\includegraphics[width=.85\columnwidth]{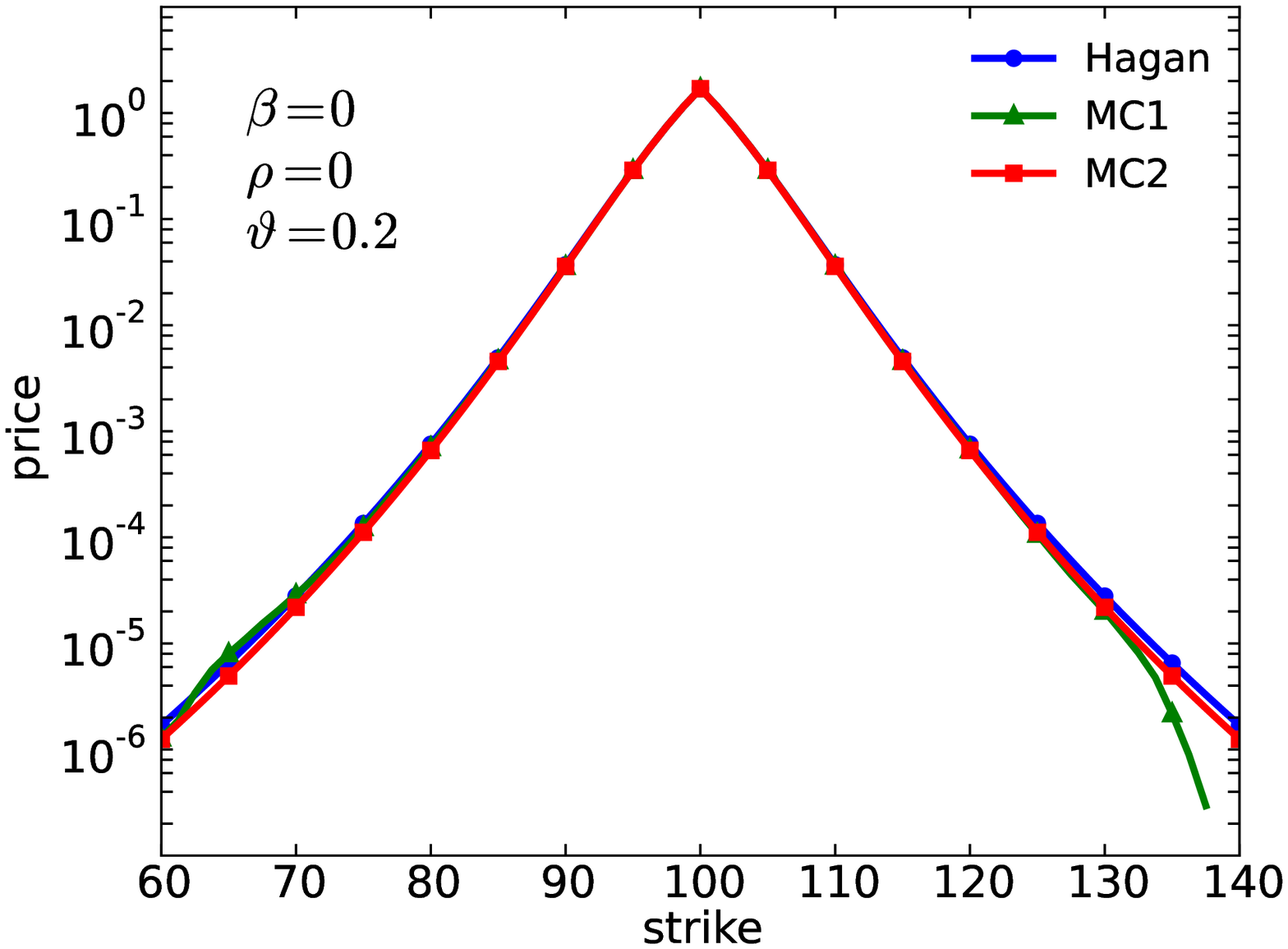}
\includegraphics[width=.85\columnwidth]{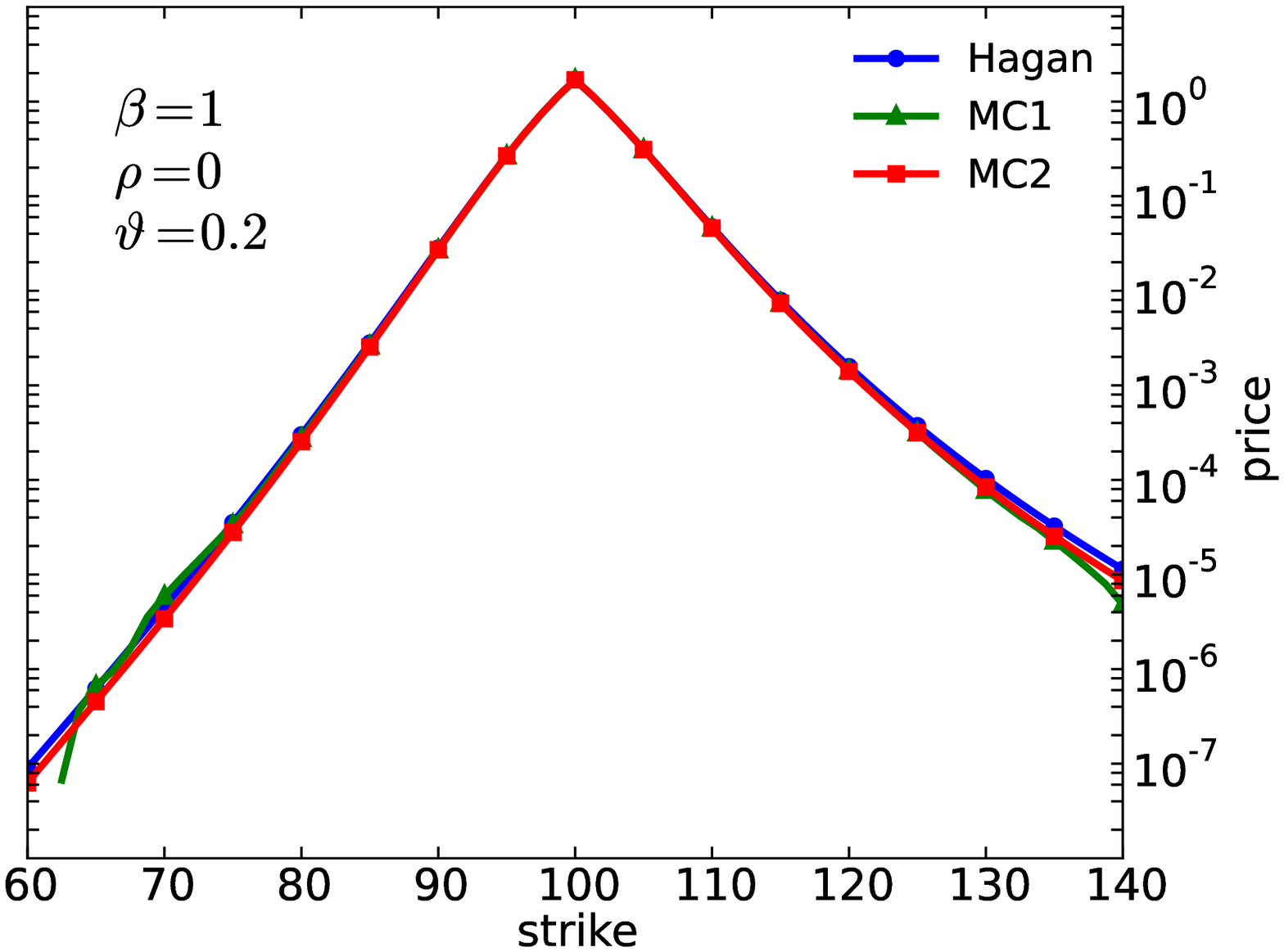}
\includegraphics[width=.85\columnwidth]{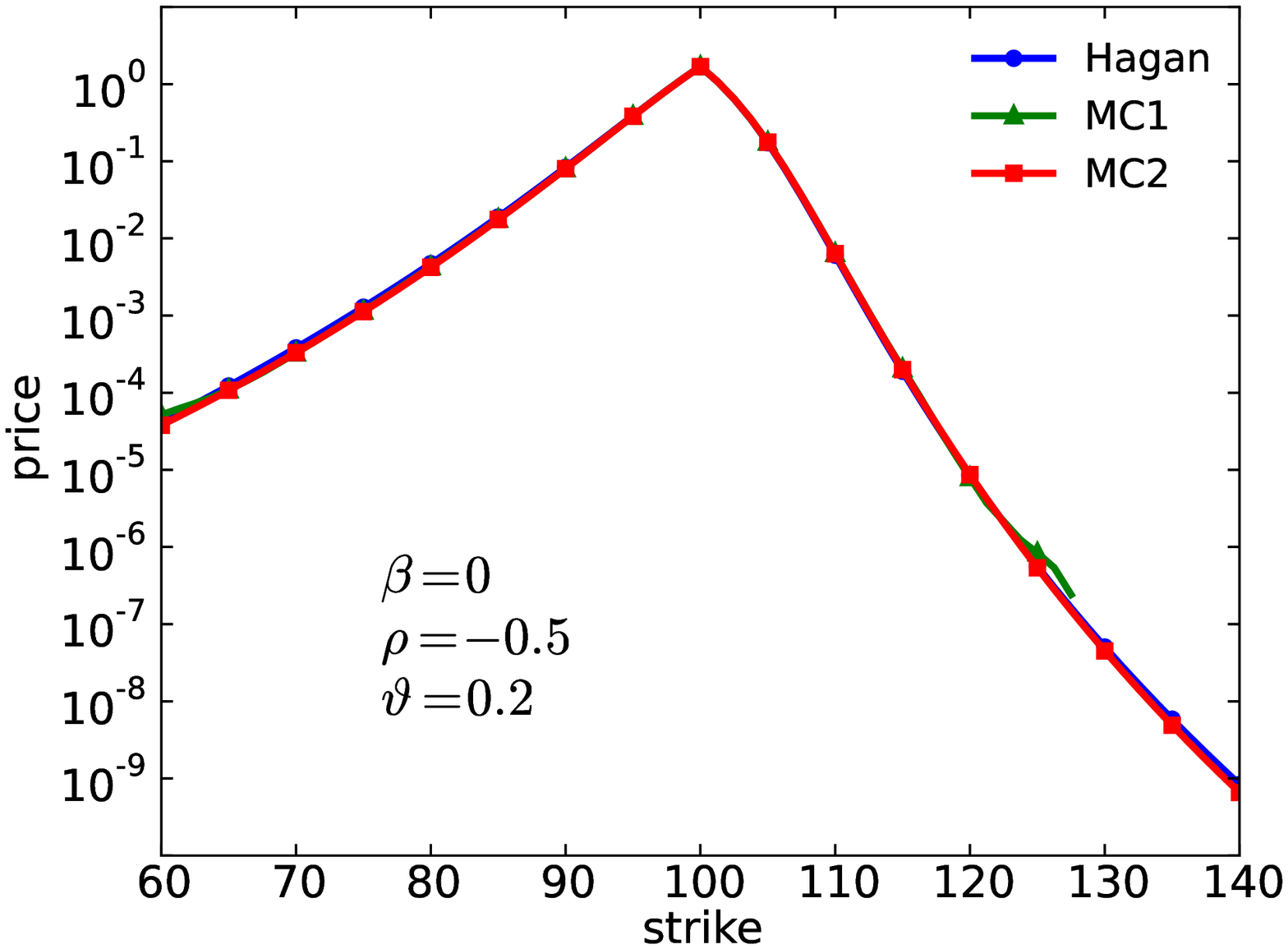}
\includegraphics[width=.85\columnwidth]{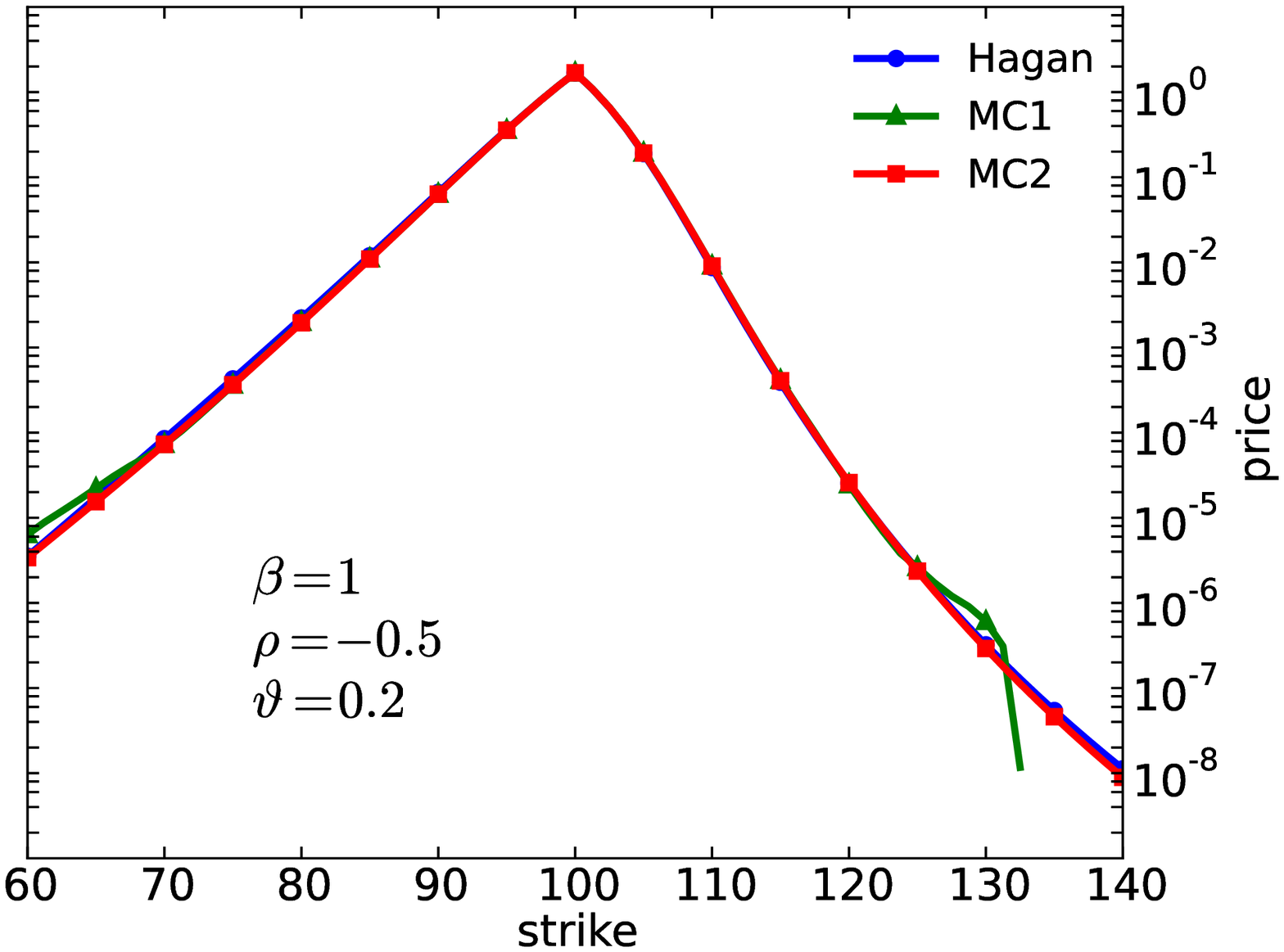}
\includegraphics[width=.85\columnwidth]{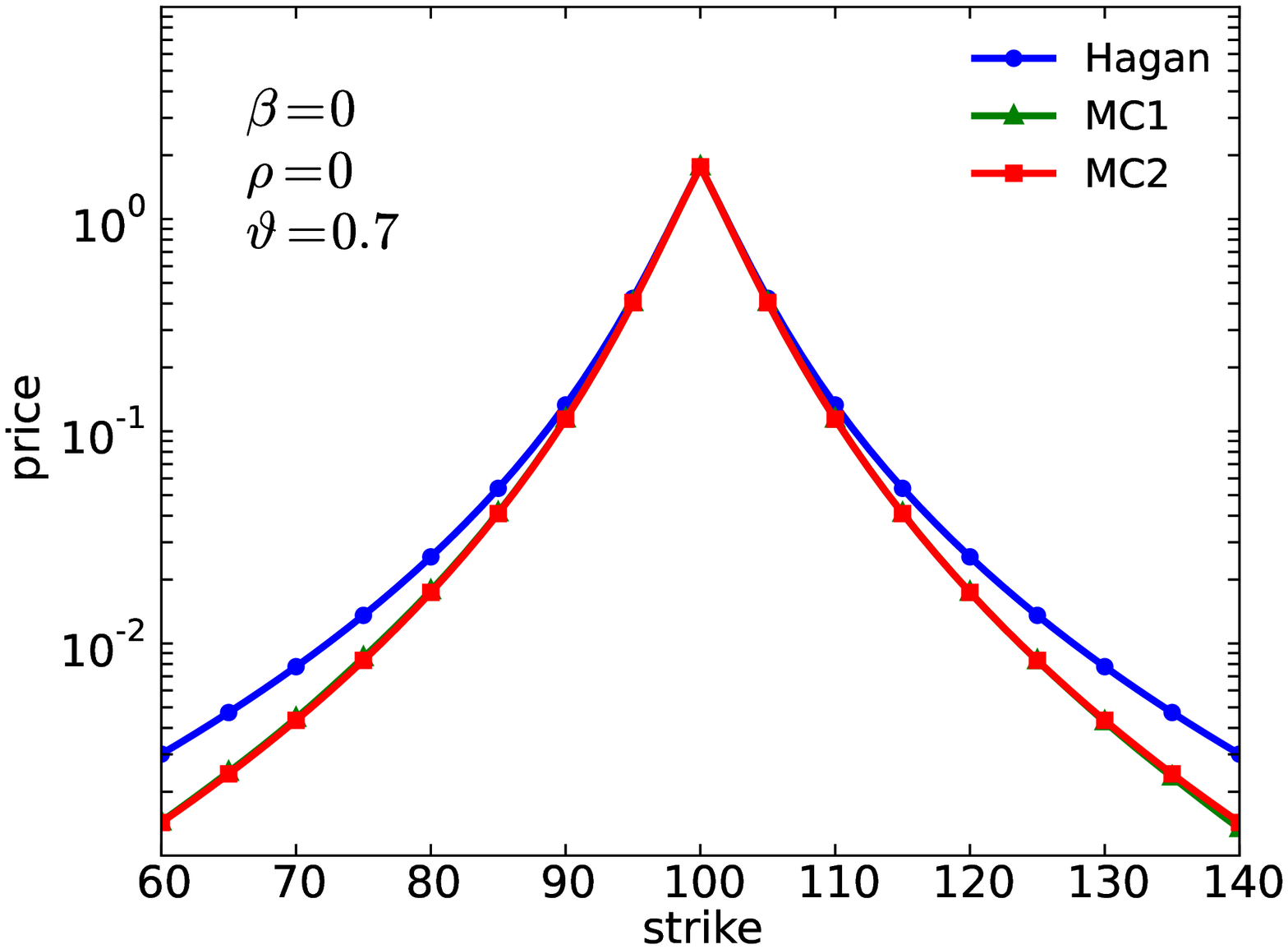}
\includegraphics[width=.85\columnwidth]{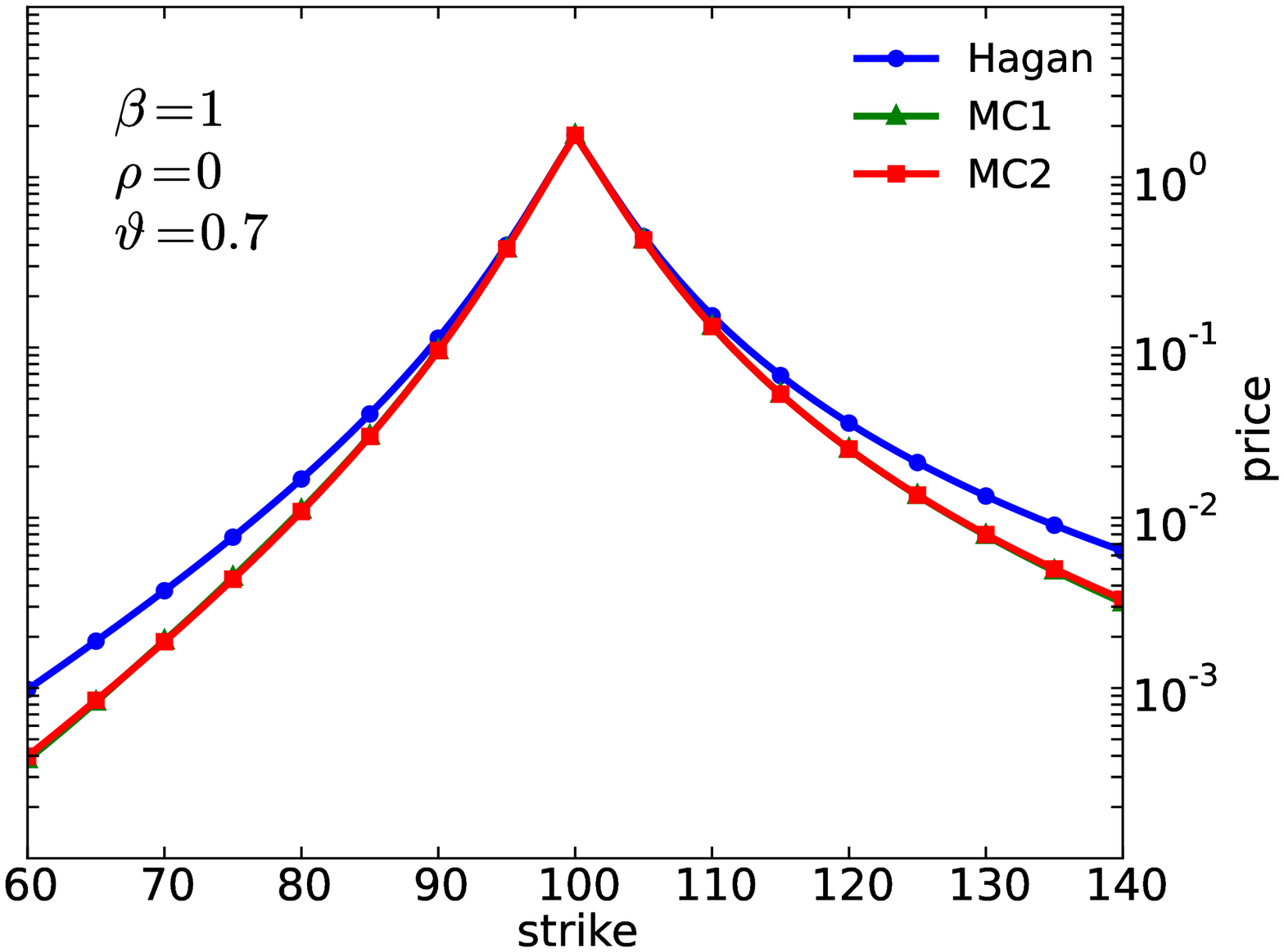}
\includegraphics[width=.85\columnwidth]{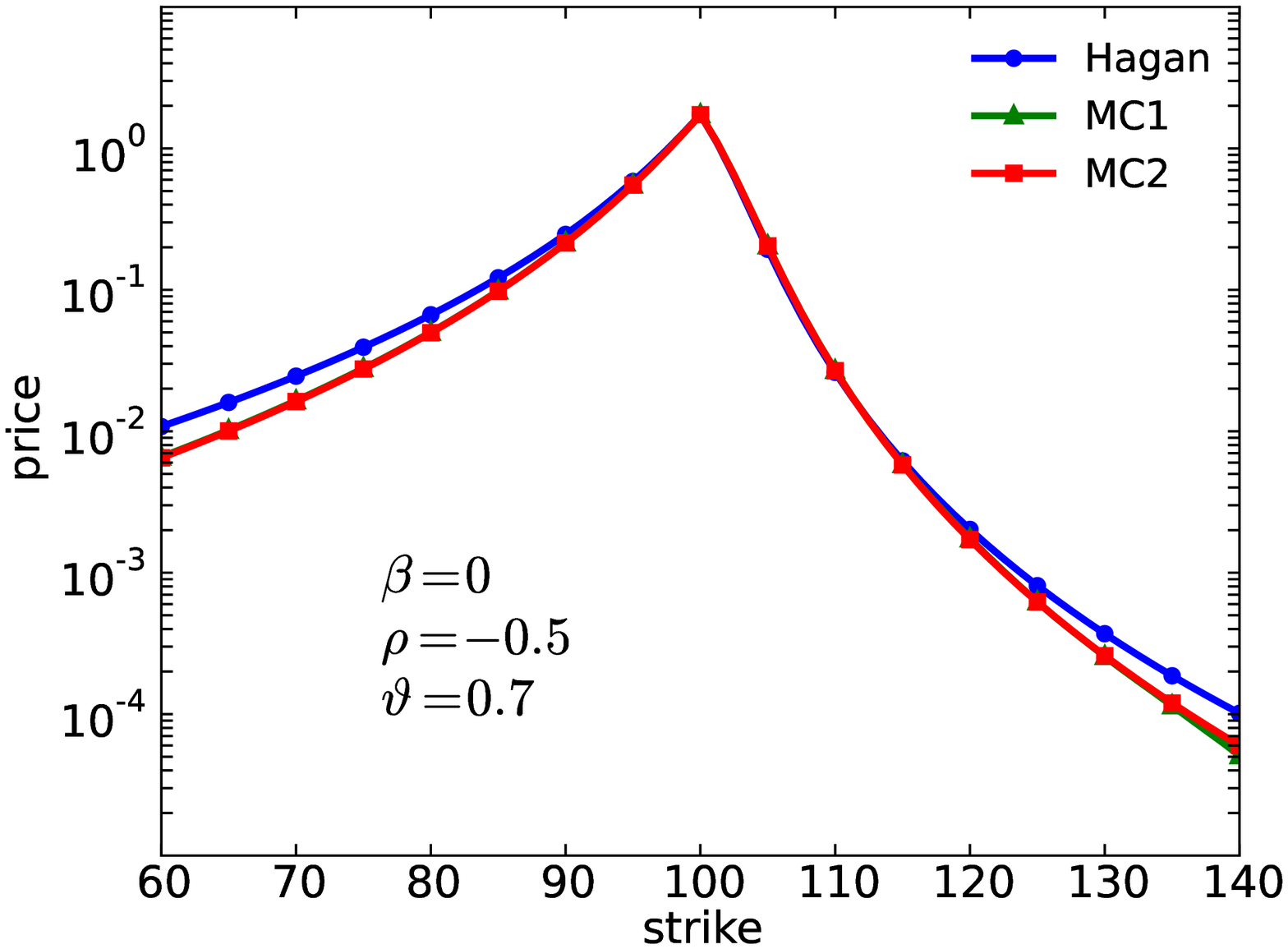}
\includegraphics[width=.85\columnwidth]{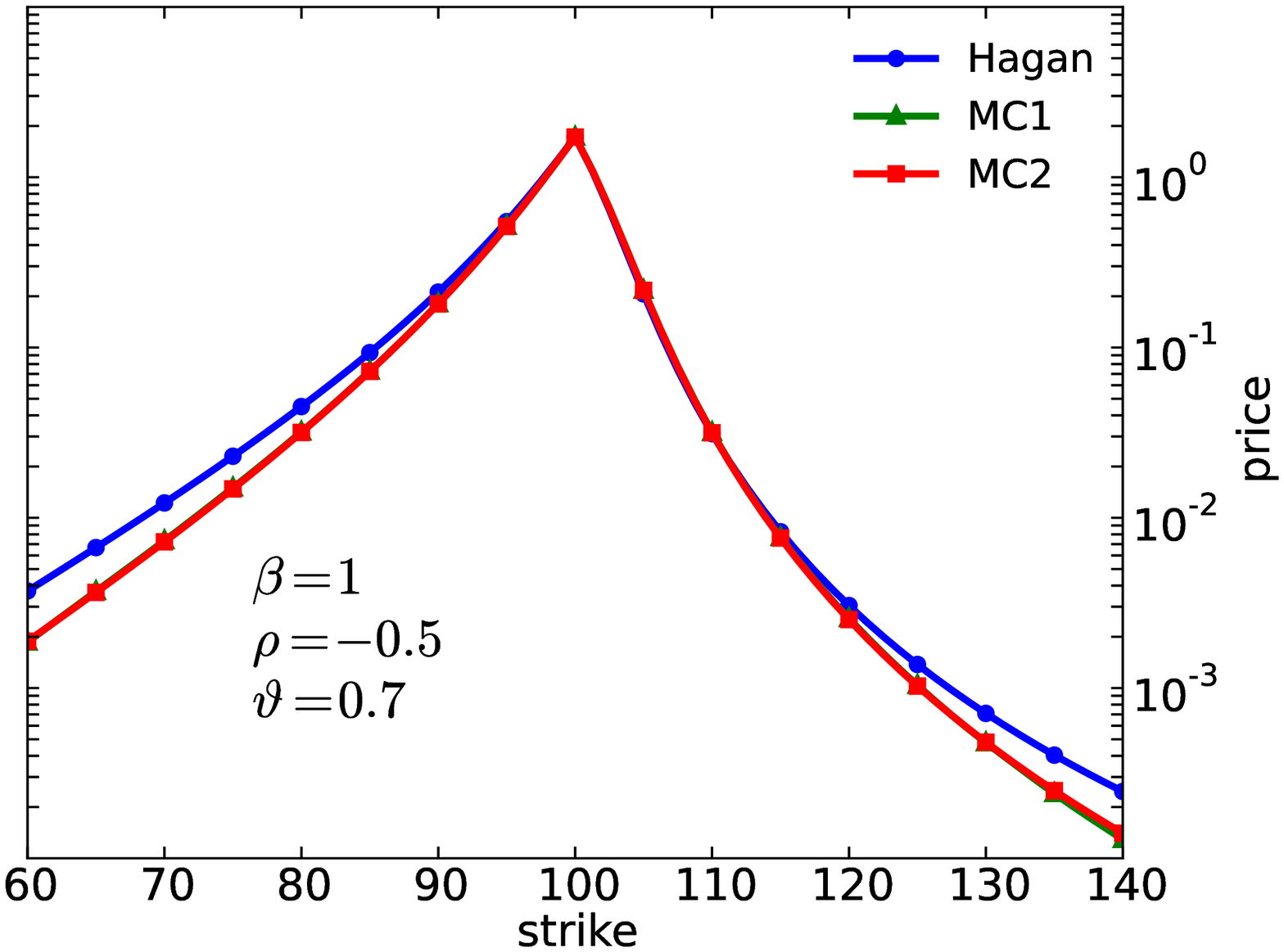}
\caption{\label{comp}Out-of-the-money Vanilla options obtained by Hagan et al. asymptotic solution (Hagan), Monte-Carlo 1 (MC1) and 2 (MC2). }
\end{figure*}

For constant $\nu$ we get $\vartheta=\nu^{2}T$. The solutions of SABR equations obtained via the two Monte-Carlo methods MC1 and MC2 are compared on FIG.\ref{comp}. As we see, for $\vartheta=0.2$ the Hagan solution has a good match for realistic option prices. For $\vartheta=0.7$ the mismatch for very out-of-the-money options is significant. From a practical point of view, the discrepancies above $10^{-2}$ matter as the price of an option is quoted up to the cents.

The same conclusion can be drawn when the correlation ($\rho$) is different from 0.


\section{Pricing Asian option using SABR $\beta=0$ equations}

Most of the literature reviewed on Asian options treats the case of the Black-Scholes model. It is a well-known fact that by using this dynamic, we assume a rigid skew that cannot be adjusted to the market. To accommodate for a wider range of skew and smile a framework such as the SABR model needs to be considered. We focus in this section on the $\beta=0$ SABR model.

For the arithmetic Asian options , under the Bachelier model, we can use the sum of Gaussian variables to characterize its average of the underlying spot price. The formula for an Asian option based on monthly prices and yearly averaging, is equivalent to Vanilla form but the volatility is now adapted ($\beta=0$, $\nu=0$, $\rho=0$).

\begin{align} 
\tilde{S}_{Average}^{Bachelier} &= S_{0}\big\{1+\frac{\sigma_{0}}{12}(\tilde{f}_{t_{1}}^{z}+\tilde{f}_{t_{2}}^{z}+\dots+ \tilde{f}_{t_{12}}^{z})\big\} \\
 &= S_{0}\Big\{1+\frac{\sigma_{0}}{12}\sum^{j=12}_{j=1}(n-j+1)[\tilde{W}_{t_{j}}-\tilde{W}_{t_{j-1}}]\Big\} \label{equAsian_bach}\\
 &= S_{0}\{1+\sigma_{A}\tilde{Q}_{B}\} \\
\sigma_{A} &= \sigma_{0}\sqrt{\frac{1}{12}\sum^{j=12}_{j=1}\Big(\frac{n-j+1}{12}\Big)^{2}} \cong \frac{\sigma_{0}}{\sqrt{3}}
\end{align}

Eventually:

\begin{equation}
C_{Asian}^{Bachelier}(F,K,\sigma_{0},\Delta T = 1) = C_{Vanilla}^{Bachelier}(F,K,\sigma_{A}, \Delta T = 1 )
\end{equation}

For the SABR $\beta=0$ two adjustments need to be done: the volatility $\tilde{\sigma}_{t}$ is variable and we have $\tilde{f}_{t}^{y}$ terms shifting each compounding Gaussian distribution. Yet we can easily generate a set of volatility trajectories and solve the Bachelier problem with variable volatility. Eventually we are able to implement the semi-analytic Monte-Carlo calculation.

\begin{align}
\tilde{S}_{Asian}^{SABR} &= S_{0}\Big\{1+\frac{1}{12}\sum_{j=1}^{j=12}(\tilde{f}_{t_{j}}^{y}+\tilde{f}_{t_{j}}^{z})\Big\} \\
&= S_{0}\Big\{1+\frac{1}{12}\sum_{j=1}^{j=12}\frac{\rho}{\nu}[\tilde{\sigma}_{t_{j}}-\sigma_{0}] + \sum_{j=1}^{j=12}(n-j+1)d\tilde{f}^{z}_{t_{j}}\Big\} \\
d\tilde{f}_{t_{j}}^{z} &= \sqrt{1-\rho^{2}}\int_{t_{j-1}}^{t_{j}}\tilde{\sigma}_{s}d\tilde{W}_{s}
\end{align}

The terms related to $\tilde{\sigma}_{t_{j}}$ are following via Monte-Carlo on $\tilde{V}_{t}$. Incorporating the terms $d\tilde{f}_{t_{j}}^{z}$ requires a further time discretization of the time increments and decomposition into independent Gaussian variables (see Appendix for details).


\section{Extension to mean reverting models}

The MC2 computation can be efficiently applied for Vanilla pricing in both the Heston/Bates and Sch\"obel-Zhu models (hence we can solve both models with time dependent coefficients). As discussed in the Appendix, the option price has the form of a straightforward average over Black-Scholes prices. The various method of simulating the Heston dynamics is presented in \cite{AndLei}.

To resolve pricing of Asian options, we need to consider a $\beta=0$ generalization of these models. In such a case the stochastic equations are ($\tilde{w}_{t=0}=w_{0},\beta=0$):

\begin{align}
d\tilde{S}_{t} &= S_{0}\Big(\frac{\tilde{S}_{t}}{S_{0}}\Big)^{\beta}\Big(\sqrt{\tilde{w}_{t}}\{\rho d\tilde{V}_{t} + \sqrt{1-\rho^{2}}d\tilde{W}_{t}\}+d\tilde{J}_{t} + \mu dt\Big)\\
d\tilde{w}_{t} &= \kappa(w_{L}-\tilde{w}_{t})dt + \nu\sqrt{\tilde{w}_{t}}d\tilde{V}_{t}
\end{align}

where $\tilde{J}_{t}$ is a compounded Poisson jump process and $\mu$ is the correcting drift factor.

Another similat model is reltaed to the Sch\"obel-Zhu model ($\zeta_{t=0}=\sigma_{0},\beta=0$)

\begin{align}
d\tilde{S}_{t} &= S_{0}\Big(\frac{\tilde{S}_{t}}{S_{0}}\Big)^{\beta}\Big(\tilde{\zeta}_{t}\{\rho d\tilde{V}_{t} + \sqrt{1-\rho^{2}}d\tilde{W}_{t}\}+d\tilde{J}_{t} + \mu dt\Big)\\
d\tilde{\zeta}_{t} &= \kappa(\zeta_{L}-\tilde{\zeta}_{t})dt + \nu d\tilde{V}_{t}
\end{align}

The models discussed above allow MC2 integration for Asian option pricing. While technically stochastic integration is straightforward, the choice and particular applicability of these models to fitting market option prices is not clear for the Authors at the moment of writing this document.

\section{Conclusion}

In this paper we have discussed a methodology for pricing options in frameworks similar to the SABR and Heston models. These dynamics are characterized by a linear secondary stochastic driver. This linearity coupled with the $\beta$ parameter allows to switch between  a Bachelier or Black-Scholes model. Alternatively, these models can base on a Shifted Log-Normal basis (``Shifted Log-Normal Backbone").

The calculation methodology is based on integrating analytically on a set of stochastic variables while the second set is integrated numerically. This methodology is known in the literature as Conditional Monte-Carlo while we call it Semi-Analytic Monte-Carlo or MC2.

The MC2 scheme can be applied to calculate option prices for all SABR/Heston models including those with time-dependent coefficients. These solutions are somehow better organized in the sense that only one variable is simulated and it relies on closed-form formulae (either Bachelier or Black-Scholes solution depending on $\beta$). The semi-analytic procedure has also a better convergence than basic Monte-Carlo schemes.

The semi-analytic MC2 calculating Vanilla option prices beyond the scope of the Hagan et al. asymptotic solutions. Separately, we can price arithmetic Asian options on markets with skew and smile using the $\beta=0$ SABR. The latter seems to be an advantage over standard calculations performed within the scope of the Black-Scholes model.

The MC2 scheme is limited to Vanilla and weakly exotic options. Its applicability is limited to $\rho=0$ for the pricing of option having a barrier condition.


\appendix{}

\section{Numerical MC2 calculations}

\subsection*{Vanilla pricing for SABR $\beta=0$}

For the SABR model we can solve explicitly the trajectory of the volatility variable functional on the path of the Brownian motion. For the SABR $\beta=0$ we get:

\begin{align}
d\tilde{S}_{t} &= S_{0}\bar{\sigma}_{t}\{\rho d\bar{V}_{t} + \sqrt{1-\rho^{2}}d\tilde{W}_{t}\} \label{equSABR_b0_app}\\
\bar{\sigma}_{t} &= \sigma_{0}\exp[\nu\bar{V}_{t}-\frac{\nu^{2}}{2}t] \label{equVol_app}
\end{align}

The solution of Eq. \ref{equSABR_b0_app} can be partially integrated and written as:

\begin{equation}
\tilde{S}_{t} = S_{0}\Big\{1+\frac{\rho}{\nu}[\bar{\sigma}_{t}-\sigma_{0}] + \sigma_{0}\sqrt{1-\rho^{2}}\int_{0}^{t}\bar{\sigma}_{s}d\tilde{W}_{s}\Big\}
\end{equation}

To calculate the second part we proceed with a time discretization $t_{n}=n\Delta t$ defining the two Brownian motions functional on two sets of independent incremental uni Gaussian variables ($\mathrm{E}[\bar{g}_{j}^{v}\tilde{g}_{j}^{w}]=\delta_{ji}\delta_{vw}$):

\begin{align}
\bar{V}_{t_{n}} = \bar{V}_{n} = \sqrt{\Delta t}\sum_{j=1}^{j=n}\bar{g}_{j}^{v} \\
\tilde{W}_{t_{n}} = \tilde{W}_{n} = \sqrt{\Delta t}\sum_{j=1}^{j=n}\tilde{g}_{j}^{w}
\end{align}

We generate a vector of Gaussian random numbers $\bar{g}_{j}^{v}$ and this gives us a trajectory for the volatility given by Eq. \ref{equVol_app}. For such a time dependent volatility we can solve the Bachelier diffusion problem as:

\begin{equation}
\tilde{S}_{n} = S_{0}\Big\{1+\frac{\sigma_{0}\rho}{\nu}[\bar{\sigma}_{n}-\sigma_{0}] + \sqrt{\Delta T}\sqrt{1-\rho^{2}}\sum_{j=1}^{j=n}\tilde{g}_{j}^{w}\bar{\sigma}_{j}\Big\}
\end{equation}

The above can be re-written in the form of a sum of independent Gaussian variables:

\begin{align}
\frac{\tilde{S}_{n}}{S_{0}} &= 1 + A + \sqrt{\Delta T}\sqrt{1-\rho^{2}}\sum_{j=1}^{j=n}\tilde{g}_{j}^{w}\tilde{\sigma}_{j}^{2} \\
\frac{\tilde{S}_{n}}{S_{0}} &= 1 + A + \sqrt{\Delta T}B\tilde{Q}_{W}
\end{align}

where 

\begin{align}
A &= \frac{\rho}{\nu}[\bar{\sigma}_{n}-\sigma_{0}] \\
B &= \sqrt{1-\rho^{2}}\sqrt{\sum_{j=1}^{j=n}\bar{\sigma}^{2}_{j}}
\end{align}

We recovered the solution of the Bachelier process but with a variable volatility ($B$ term) and a corrected $S_{0}$ ($A$ term). The above allows calculating the Vanilla price for a selected $m^{th}$ trajectory:

\begin{equation}
C_{m}^{Num}(F,K,\sigma_{0},\Delta T) = C_{V}^{Bachelier}(F+A_{m},K,B_{m},\Delta T)
\end{equation}

Eventually we average over the \textit{M} Monte-Carlo sampled volatility trajectories:

\begin{align}
&C_{V}^{SABR\ \beta=0}(F,K,\sigma_{0},\Delta T) = \nonumber\\ &\frac{1}{M}\sum_{m=1}^{m=M}C_{V}^{Bachelier}(F+A_{m},K,B_{m},\Delta T)
\end{align}

The price of a Vanilla call under the Bachelier model is as follows:

\begin{align}
C_{V}^{B}(F,K,\sigma_{0},\Delta T) &= (F-K)\mathcal{N}[d_{0}] + \frac{F\sigma_{0}\sqrt{\Delta T}}{\sqrt{2\pi}}e^{-\frac{1}{2}d_{0}^{2}} \\
d_{0} &= \frac{K-F}{F\sigma_{0}\sqrt{\Delta T}}
\end{align}

\subsection*{Vanilla pricing for SABR $\beta=1$}

For the SABR $\beta=1$ we get:

\begin{align}
d\tilde{S}_{t} &= \tilde{S}_{t}\bar{\sigma}_{t}\{\rho d\bar{V}_{t} + \sqrt{1-\rho^{2}}d\tilde{W}_{t}\} \label{equSABR_b1_app}\\
\bar{\sigma}_{t} &= \sigma_{0}\exp[\nu\bar{V}_{t}-\frac{\nu^{2}}{2}t] \label{equVol1_app}
\end{align}

We proceed along similar lines and solve partially Eq. \ref{equSABR_b1_app}:

\begin{align}
\begin{split}
\tilde{S}_{t} = S_{0}\exp\Big\{&\frac{\rho}{\nu}[\bar{\sigma}_{t}-\sigma_{0}] \\&- \frac{1}{2}\int_{0}^{t}\bar{\sigma}_{s}^{2}ds \\&+ \sqrt{1-\rho^{2}}\int_{0}^{t}\bar{\sigma}_{s}d\tilde{W}_{s} \Big\}
\end{split} \\
\begin{split}
= S_{0}\exp\Big\{&\frac{\rho}{\nu}[\bar{\sigma}_{t}-\sigma_{0}] -\frac{1}{2}\rho^{2}\int_{0}^{t}\bar{\sigma}_{s}^{2}ds\\&- \frac{1}{2}(1-\rho^{2})\int_{0}^{t}\bar{\sigma}_{s}^{2}ds \\&+ \sqrt{1-\rho^{2}}\int_{0}^{t}\bar{\sigma}_{s}d\tilde{W}_{s} \Big\}
\end{split}
\end{align}

We solve the integral parts of the above equation using discretization for the two Wiener trajectories. The discretization must be done avoiding lattice errors which checked by getting a Black-Scholes type formula:

\begin{equation}
\ln\Big[\frac{\tilde{S}_{n}}{S_{0}}\Big] = A^{BS} - \frac{1}{2}\Delta T B^{2} + \sqrt{\Delta T}B\tilde{Q}_{W}
\end{equation}

where 

\begin{align}
A^{BS} &= \frac{\rho}{\nu}[\bar{\sigma}_{n}-\sigma_{0}] -\frac{1}{2}\rho^{2}\sum_{j=1}^{j=n}\bar{\sigma}^{2}_{j}\sqrt{\Delta t} \\
B &= \sqrt{1-\rho^{2}}\sqrt{\sum_{j=1}^{j=n}\bar{\sigma}^{2}_{j}}
\end{align}

The Black-Scholes price for a single volatility trajectory is:

\begin{equation}
C_{m}^{Num}(F,K,\sigma_{0},\Delta T) = C_{V}^{BS}(F+A_{m}^{BS}, K,B_{m}, \Delta T)
\end{equation}

And the SABR price avergaed over the volatility trajectories is:

\begin{equation}
\begin{split}
&C_{V}^{SABR\ \beta=1}(F,K,\sigma_{0},\Delta T) =\\& \frac{1}{M}\sum_{m=1}^{m=M}C_{V}^{BS}(F+A_{m}^{BS},K,B_{m},\Delta T)
\end{split}
\end{equation}

\begin{figure}
\centering
\includegraphics[width=\columnwidth]{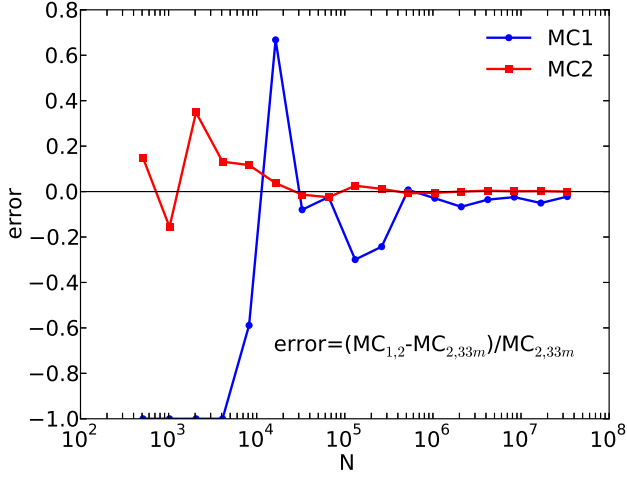}
\caption{Relative error of the Monte-Carlo methods compared to MC2 with 33 ($2^{25}$) millions trajectories  for an out-of-the-money Vanilla option (K=80\%). The MC2 error is more stable compared to MC1, even for small number of simulations.}
\end{figure}

\subsection*{Asian options for SABR $\beta=0$}

We select a trajectory for the volatility variable and follow on the result given by Eq. \ref{equAsian_bach}:

\begin{equation}
\begin{split}
\frac{\tilde{S}_{A}^{SABR}}{S_{0}} &= 1 + \frac{1}{12}\sum_{j=1}^{j=12}\bar{f}_{t_{j}}^{y} \\&+ \frac{1}{12}\sqrt{1-\rho^{2}}\sqrt{\Delta T}\sum_{j=1}^{j=12}(n-j+1)d\tilde{f}_{t_{j}}^{z}
\end{split}
\end{equation}

We calculate the shift term:

\begin{equation}
A' = \frac{1}{12}\sum_{j=1}^{j=12}\frac{\rho}{\nu}[\bar{\sigma}_{t_{j}}-\sigma_{0}]
\end{equation}

And decompose $\tilde{f}_{t_{i}}^{z}$ into elementary variables:

\begin{equation}
d\tilde{f}_{t_{j}}^{z} =\sum_{i=1}^{i=N_{j}}\tilde{g}_{i}^{w}\bar{\sigma}_{i}-\sum_{i=1}^{i=N_{j-1}}\tilde{g}_{i}^{w}\bar{\sigma}_{i}
\end{equation}

Please note that this decomposition accounts for the dynamic aspect of volatility. Eventually we calculate:

\begin{equation}
B'=\sqrt{1-\rho^{2}}\sqrt{\sum_{j=1}^{j=12}\Big(\sum_{i=1}^{N_{j}}\bar{\sigma}_{i}^{2}-
\sum_{i=1}^{N_{j-1}}\bar{\sigma}_{i}^{2}\Big)(n-j+1)^{2}}
\end{equation}
\begin{equation}
\frac{\tilde{S}_{A}^{SABR}}{S_{0}} = 1 + A' + \sqrt{\Delta T}B' \tilde{Q}_{W}
\end{equation}

This means that the price for an Asian option has the form:

\begin{equation}
\begin{split}
&C_{Asian}^{SABR\ \beta=0}(F,K,\sigma_{0},\Delta T) =\\ &\frac{1}{M}\sum_{m=1}^{m=M}C_{V}^{Bachelier}(F+A'_{m},K,B'_{m},\Delta T)
\end{split}
\end{equation}

\begin{figure}
\includegraphics[width=.99999\columnwidth]{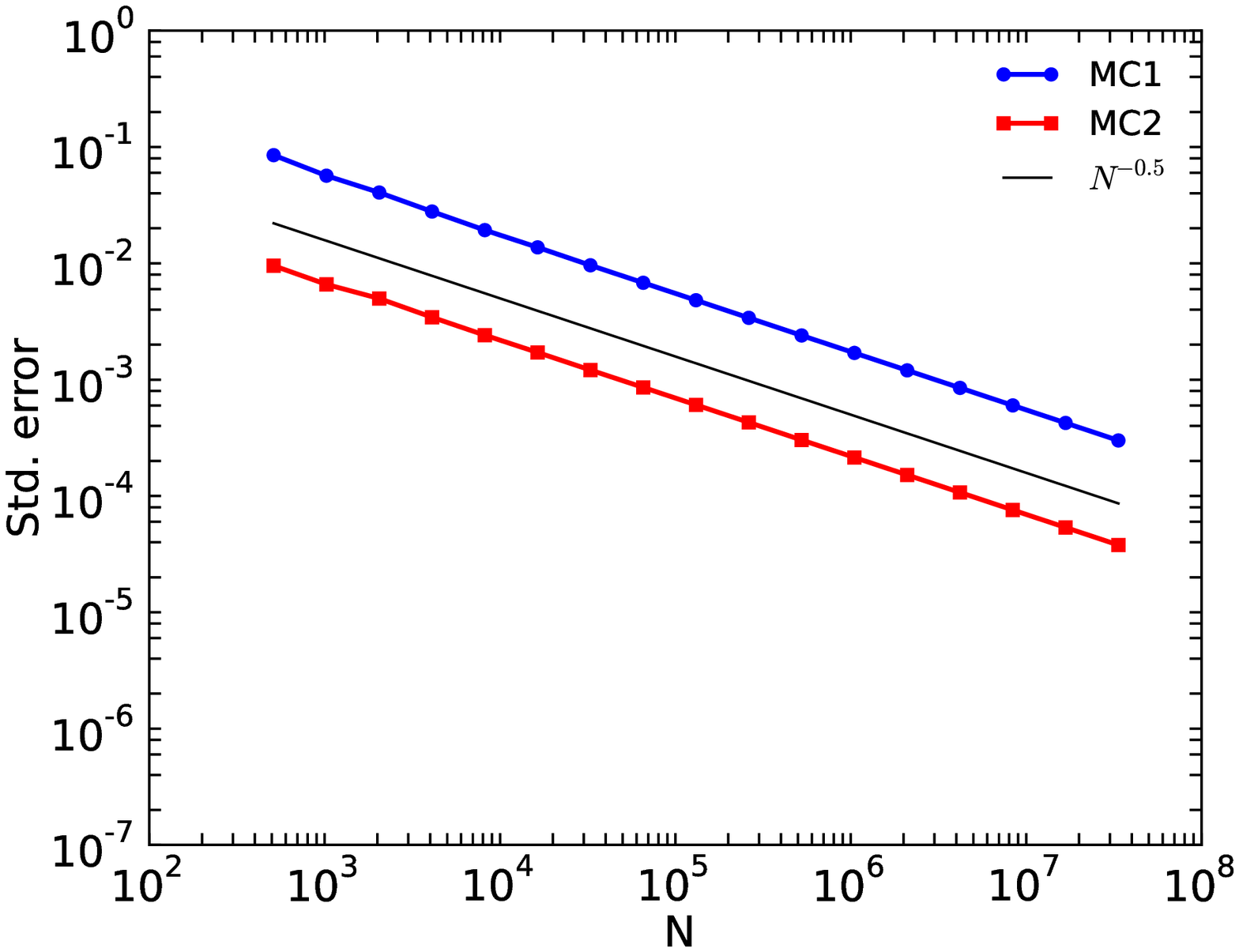}\llap{\makebox[.84\columnwidth][l]{\raisebox{.9cm}{\includegraphics[width=.43\columnwidth]{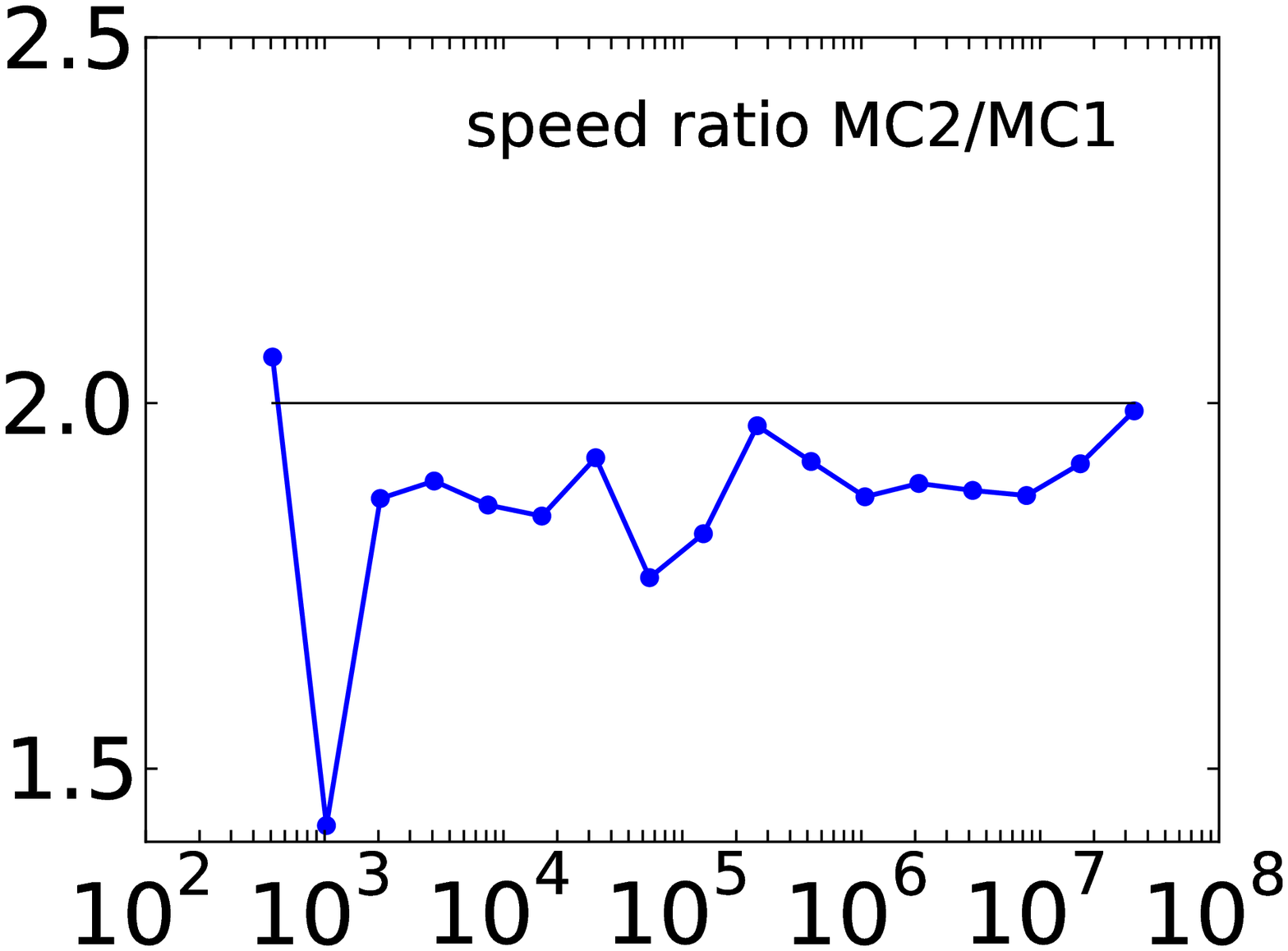}}}}
\caption{\label{Aconv}Comparison of the convergence between Monte-Carlo 1 (MC1) and 2 (MC2) for Asian option based on SABR $\beta=0$ dynamic. \textit{Inset:} MC2 is about twice as fast as MC1.}
\end{figure}

\begin{figure*}
\centering
\includegraphics[width=.85\columnwidth]{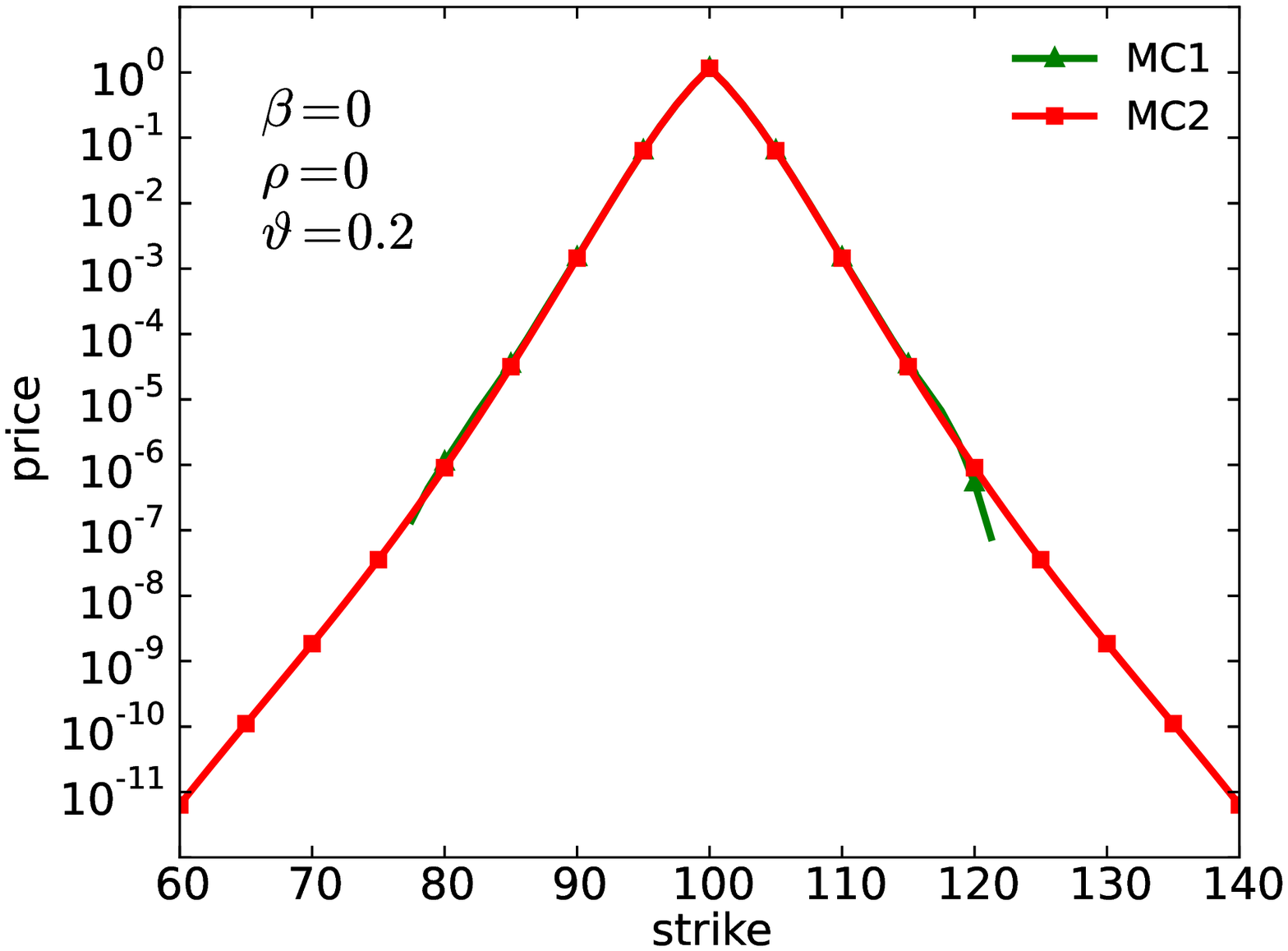}
\includegraphics[width=.85\columnwidth]{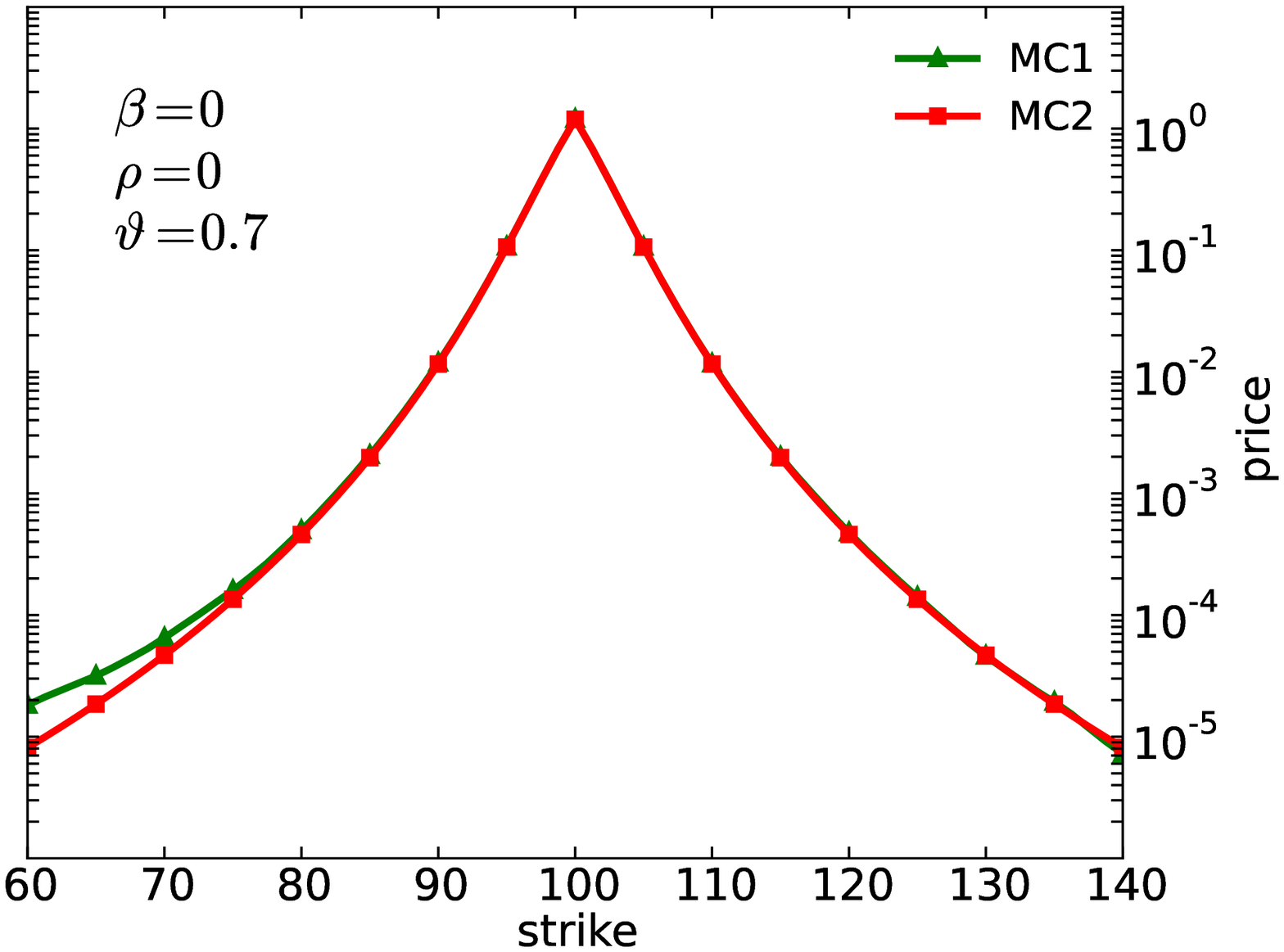}
\includegraphics[width=.85\columnwidth]{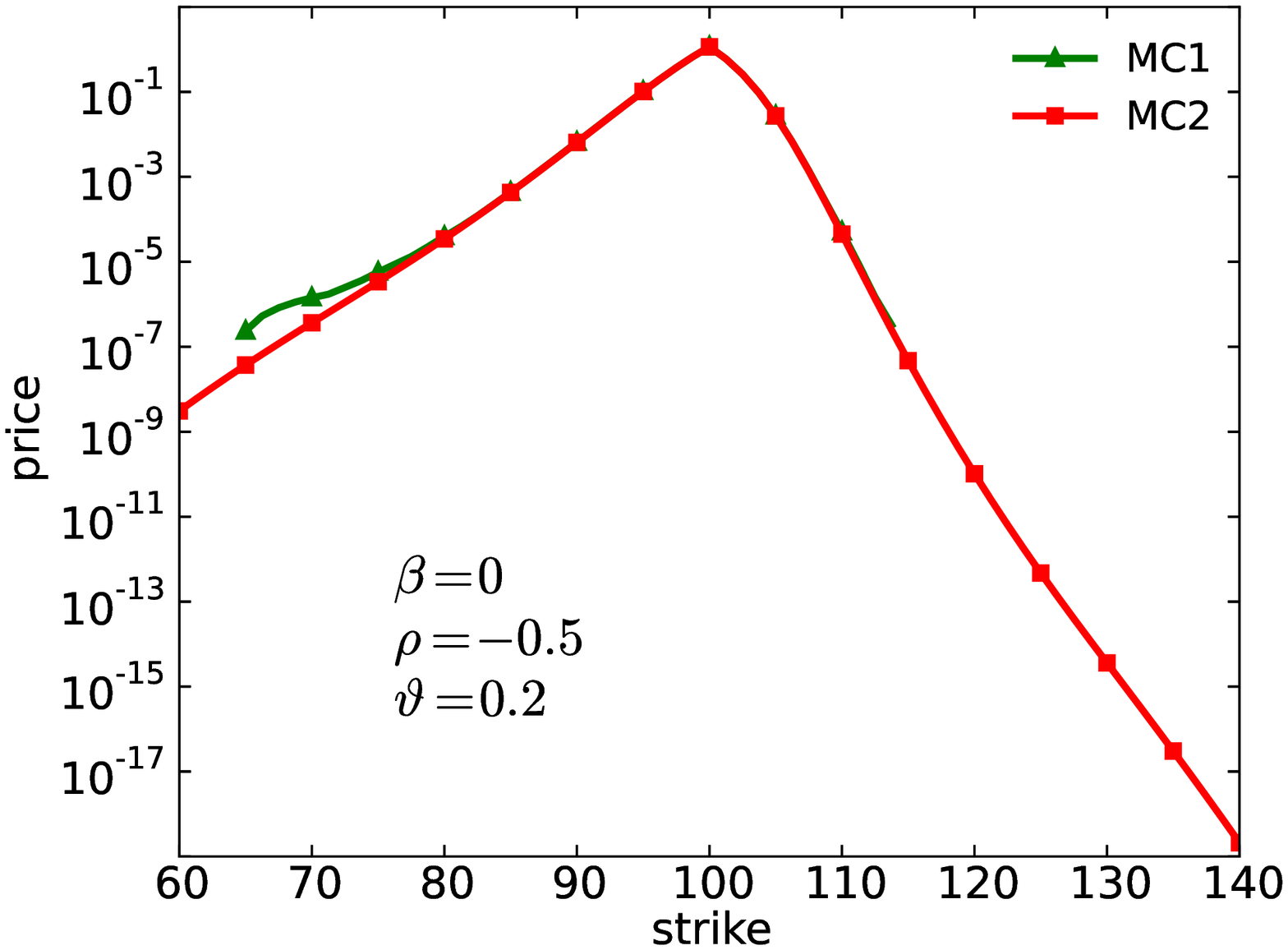}
\includegraphics[width=.85\columnwidth]{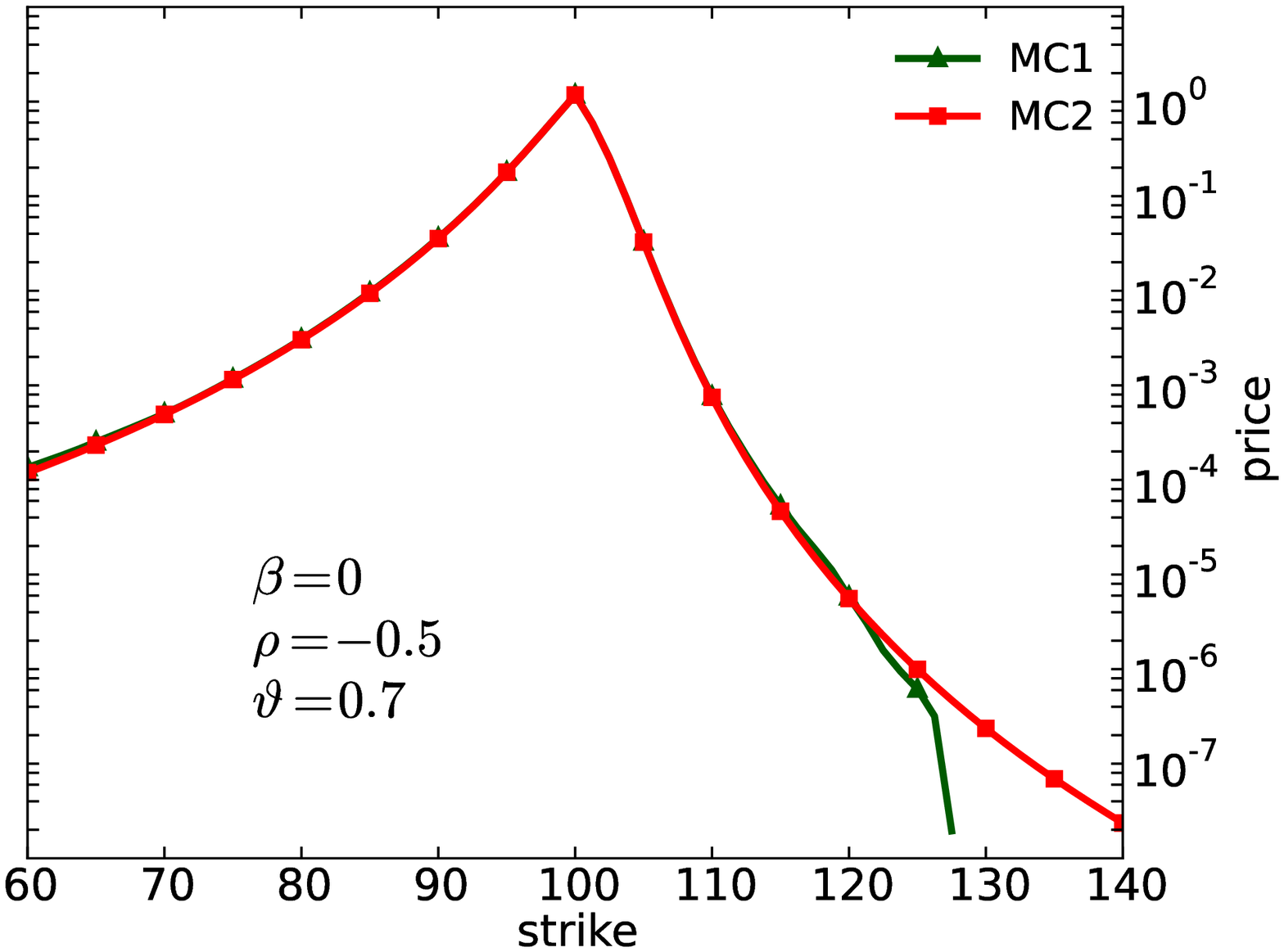}
\caption{Out-of-the money arithmetic Asian options obtained by Monte-Carlo 1 (MC1) and 2 (MC2) based on SABR $\beta=0$ dynamic. The two methods have a good match.}
\end{figure*}

\clearpage


\nocite{Duf2, Duf3,JonPooRoc, Kle, Pra, Tsa, Tsa2}
\bibliography{SABR_semi_analytic_MC}{}
\bibliographystyle{plain}


\end{document}